\documentstyle[12pt,epsf]{article}
\begin{document}
\title{Chargino contribution to the rare decay $b\to ss\bar{d}$}
\author{Xiao-Hong Wu\footnote{wuxh@th.phy.pku.edu.cn}
and Da-Xin Zhang\footnote{dxzhang@mail.phy.pku.edu.cn}\\
\it\small Institute of Theoretical Physics,\\[-3mm]
\it\small School of Physics, Peking University, Beijing 100871,
China}
\date{February 6, 2004}

\maketitle

\begin{abstract}
The rare decay $b\to ss\bar{d}$ is studied in the supersymmetric
standard model by considering the contribution from the chargino
box diagrams.
We find that this contribution amounts to $10^{-9}$ in the
branching ratio.
\end{abstract}

\vspace{0.5cm} {\it PACS}: 12.15.Mm, 13.90.+i, 14.40.Lb, 14.40.Nd

% 12.15.Ji, 13.25.Hw, 12.60.Jv

\newpage
A very important aspect in the study of  B physics is to expose
possible virtual effects  due to physics beyond the Standard
Model(SM). These virtual effects are likely to be hidden in the
processes induced by flavor changing neutral current (FCNC)
interactions, since the SM contributions come from loop diagrams
so that they do not always dominate over the new physics contributions.
Furthermore, since the SM contributions serve as the background,
the rare decays with unobservable SM contributions are more
suitable to probe the new physics signals. As was pointed out in
\cite{hlsz} and \cite{hlsz2}, the rare decay $b\to s s\bar d$ is a
clean channel with less SM background. The quark-level transition
materializes as decays of B mesons into $S=2$ final states, of
which a fraction $1/4$ are states with two charged kaons which can
be identified at future experiments. Several exclusive channels of
$b\to s s\bar d$ transition have been searched by different
groups~\cite{data}. More data will improve the bounds on these transitions and
on the possible new physics.

The rare decay $b\to s s\bar d$ in the SM proceeds with a box
diagram which is strongly suppresses by the second order weak
interactions and by the Glashow-Maiani-Illiopoulous (GIM)
cancellations. In the SM this branching ratio is too small to be
observable. This may help to expose the possible signal of new
physics clearly from the background of the SM. Motivated by this
observation, several new physics models has been studied for this
process\cite{hlsz,hlsz2,singer}.

In this Letter we will consider the chargino contribution to
$b\rightarrow s s \bar{d}$ within the Minimal Supersymmetric
Standard Model (MSSM). In the MSSM, the charged Higgs contribution
is negligible\cite{hlsz2}. The gluino-squark contribution has been
studied in \cite{hlsz} which can be as large as $10^{-8}$ in the
branching ratio, depending on the parameter space. The chargino
contribution, which is usually another important source of the
FCNC interactions, is to be studied below.

In the MSSM, the chargino contribution to $b\rightarrow s s
\bar{d}$ is from the chargino and the up-type squark diagrams.
These up-type squarks are not degenerate and do not align with the
up-type quarks in the super-multiplets. We use the mass insertion
approach to estimate the chargino contribution. Following the
notations of \cite{buras98}, the couplings of the left-handed
down-type quarks, the left-handed up-type squarks and the
charginos are flavor diagonal. In this basis, the mass-squared
matrix for the up-type squarks is not diagonal. The
off-diagonal elements,  which will be denoted as $\Delta^{LL}_{ij}$
($i,j=1,2,3$), are taken as the two-point vertices. Consequently,
the insertions of these two-point vertices on the up-type squarks
propagators inside the loops induce the transitions between
different external down-type quarks. Similarly, since the
right-handed stop couples with the higgsino components of the
charginos with a strength which is proportional to large Yukawa
coupling of the top quark, there are also the diagrams with the
insertions of the left-handed up-type squarks
and the right-handed stop $\Delta^{LR}_{i3}$ ($i=1,2,3$) inside the loops.
Note that the possible large splitting in
the masses of the left-handed and the right-handed stops induces
another source of FCNC. The relevant effects are parameterized
as\cite{buras98}
\begin{eqnarray}
R^{LL}_{sb} &=& \frac{K^\ast_{is} K_{jb}
\Delta^{LL}_{ij}}{K^\ast_{ts}
  K_{tb} \tilde{m}^2}, \hspace{5mm}
 R^{LL}_{sd} = \frac{K^\ast_{is} K_{jd} \Delta^{LL}_{ij}}{K^\ast_{ts}
  K_{td} \tilde{m}^2}, \nonumber\\
R^{RL}_{tb} &=& \frac{K_{ib} \Delta^{RL}_{3i}}{ K_{tb}
\tilde{m}^2},
  \hspace{10mm}
 R^{LR}_{st} = \frac{K^\ast_{is} \Delta^{LR}_{i3}}{K^\ast_{ts} \tilde{m}^2},
  \nonumber\\
R^{RL}_{td} &=& \frac{K_{id} \Delta^{RL}_{3i}}{K_{td}
\tilde{m}^2},
\end{eqnarray}
where $K$ is the CKM matrix, and $\tilde{m}$ is the averaged
up-type squark mass. The repeating indices $i,j$ in the same
equation indicate sum over from $1$ to $3$. The quantities $R$'s
are constrained directly by several experimental observables, {\it
e.g.} $R^{LL}_{sb}$, $R^{LR}_{st}$ and $R^{RL}_{tb}$ are
constrained by the mass difference $\Delta M_{B_S}$ and by the
branching ratio of $b \rightarrow s \gamma$, and $R^{LL}_{sd}$,
$R^{LR}_{st}$, $R^{RL}_{td}$ are constrained by the mass
difference $\Delta M_K$.
%mass differences
%of $\bar{B}^0_s - B^0_s$, $\bar{K}^0 - K^0$ system $\Delta
%M_{B_s}$, $\Delta M_K$ and Br($b \rightarrow s \gamma$) with
%$\Delta M_{B_s}$ and Br($b \rightarrow s \gamma$) constrained
%$R^{LL}_{sb}$, $R^{LR}_{st}$ and $R^{RL}_{tb}$, and $\Delta M_K$
%constrained $R^{LL}_{sd}$, $R^{LR}_{st}$ and $R^{RL}_{td}$.

\begin{figure}
\epsfxsize=16cm \epsfysize=16cm \centerline{\epsffile{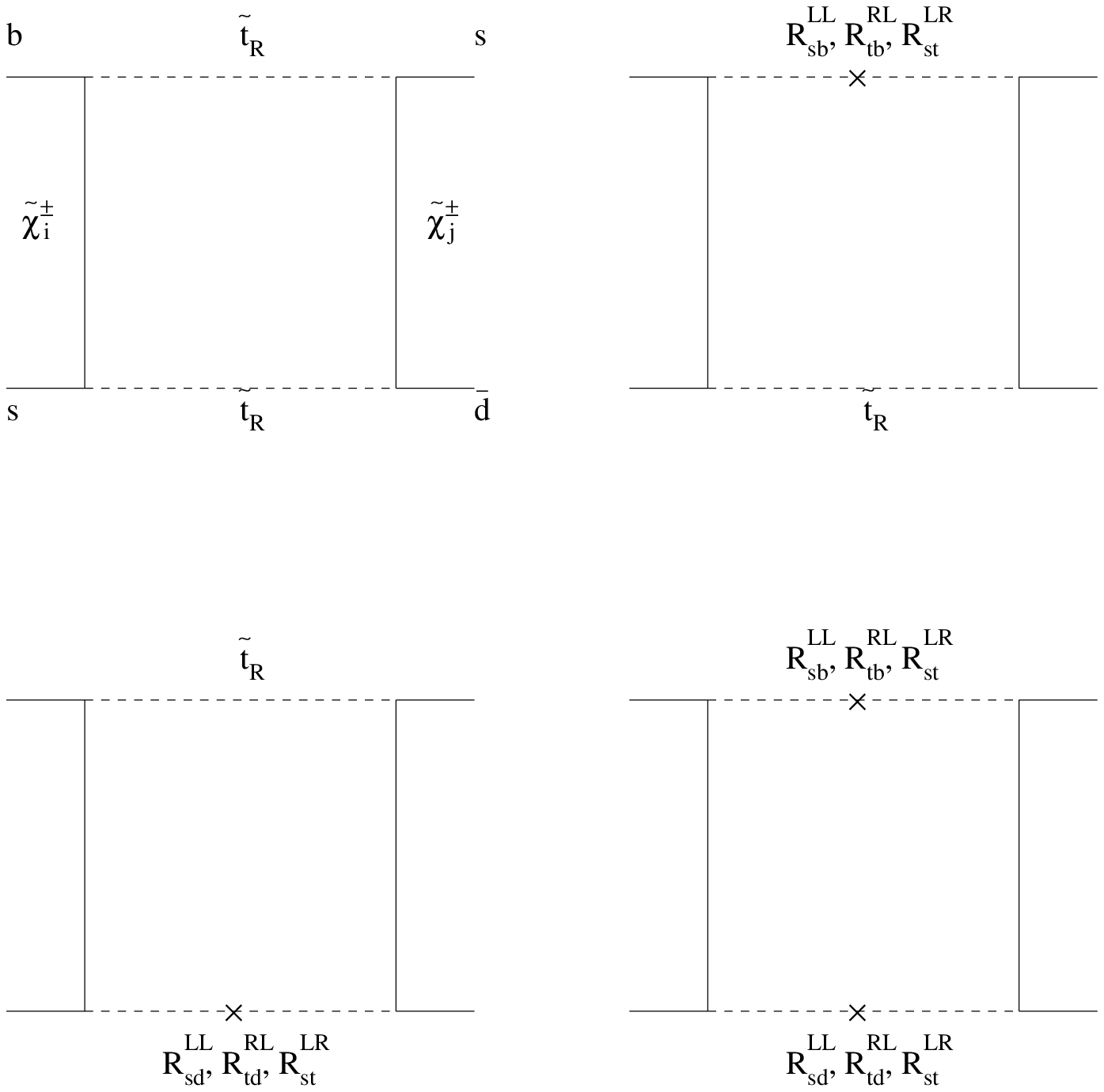}}
\caption{ The Feynman diagrams for process $b \rightarrow s s
\bar{d}$. Crossing diagrams are not shown. } \label{feyndiagram}
\end{figure}

The Feynman diagrams relevant to the process $b \rightarrow s s
\bar{d}$ are shown in {\bf Fig}.~\ref{feyndiagram}. We calculate
the decay width from the chargino contribution to  be
\begin{eqnarray}
\Gamma &=& \frac{m_b^5}{48 (2\pi)^3}
 \sum_{i,j} | \frac{g^4}{64\pi^2} K_{tb} K_{ts}^\ast K_{td} K_{ts}^\ast
   \frac{1}{m_{\tilde{\chi}_j^\pm}^2} \nonumber\\
&& [ V_{i1} V_{i1} V_{j1} V_{j1}
 x_j^2 f(x_j, x_j, x_j, x_j, x_{ij}) R^{LL}_{sb} R^{LL}_{sd}\nonumber\\
&& - h_t V_{i1} V_{i1} V_{j1} V_{j2} x_j^2 f(x_j, x_j, x_j,
x_{Rj}, x_{ij})
 (R^{RL}_{tb} R^{LL}_{sd} + R^{LR}_{st} R^{LL}_{sd} +
 R^{LL}_{sb} R^{LR}_{st} + R^{LL}_{sb} R^{RL}_{td}) \nonumber\\
&& + h^2_t ( V_{i1} V_{i2} V_{j1} V_{j2} x_j f(x_j, x_j, x_{Rj},
x_{ij})
 (R^{LL}_{sb} + R^{LL}_{sd}) \nonumber\\
&& ~~~~~~+x_j^2 f(x_j, x_j, x_{Rj}, x_{Rj}, x_{ij}) ( V_{i1}
V_{i2} V_{j1} V_{j2}
  (R^{LR}_{st} R^{LR}_{st}+ R^{RL}_{tb} R^{RL}_{td}) \nonumber\\
  &&~~~~~~~~~~~~~~~~~~~~~~~~~~~~~~~~~~~~~~~~~+ V_{i1} V_{i1} V_{j2} V_{j2}
  ( R^{RL}_{tb} R^{LR}_{st} + R^{LR}_{st} R^{RL}_{td} ) ) ) \nonumber\\
&& - h^3_t V_{i1} V_{i2} V_{j2} V_{j2} x_j f(x_j, x_{Rj}, x_{Rj},
x_{ij})
 ( R^{RL}_{tb} + 2 R^{LR}_{st} + R^{RL}_{td} ) \nonumber\\
&& + h^4_t V_{i2} V_{i2} V_{j2} V_{j2} f(x_{Rj}, x_{Rj}, x_{ij})
 ]|^2
\end{eqnarray}
with $x_j = \tilde{m}^2/m_{\tilde{\chi}_j^\pm}^2$, $x_{Rj} =
m_{\tilde{t}_R}^2/m_{\tilde{\chi}_j^\pm}^2$, $x_{ij} =
m_{\tilde{\chi}_i^\pm}^2/m_{\tilde{\chi}_j^\pm}^2$, $V$ is the
matrix to diagonalize the chargino mass matrix $X_{\rm chargino}$
with $U X_{\rm chargino} V^{-1} = {\rm
diag}(m_{\tilde{\chi}^\pm_1}, m_{\tilde{\chi}^\pm_2})$,
and the Yukawa couplings are defined as $h_t = m_t/(\sqrt{2} m_W \sin\beta)$,
$h_b = m_b/(\sqrt{2} m_W \cos\beta)$.
The loop functions are~\cite{buras98}
\begin{eqnarray}
 f(x,y,z) &=& x^2 \ln x/( (x-y)
(x-z)(x-1) ) + y^2 \ln y/( (y-x) (y-z)(y-1) )\nonumber\\
 &&+ z^2\ln z/( (z-x)
(z-y)(z-1) ), \nonumber\\
f(x,y,p,q) &=& [ f(x,p,q) - f(y,p,q) ]/(x - y), \nonumber\\
f(x,y,p,q,m) &=& [ f(x,p,q,m) - f(y,p,q,m) ]/(x - y).
\end{eqnarray}
We would like to mention that the chargino contribution of the box
diagrams is proportional to $1/m_{\tilde{\chi}^\pm_j}^2$. If both
charginos are heavy, their contribution to Br($b \rightarrow s s
\bar{d}$) is small, so the main contribution comes from the
lighter chargino $\tilde{\chi}^\pm_1$ with a small mass (say $<
200$GeV).

Numerically we take into account the constraints of $\Delta
M_{B_S} \ge 14.4{\rm ps}^{-1}$~\cite{sto}, $2\times 10^{-4} \le
{\rm Br}(B\rightarrow X_s \gamma) \le 4.5\times 10^{-4}$, and the
lower bounds on the superparticles~\cite{pdg}. As for the
constraint of $\Delta M_K$, we demand the chargino contribution to
$\Delta M_K$ does not exceed the experimental value of $\Delta
M_K^{\rm exp}=(3.489 \pm 0.008) \times 10^{-15}$GeV.

The calculation of this process contains 14 parameters, $6$ of
them are $\Delta^{LL}_{ij}$ with $ij=11,12,13,22,23,33$ which
contribute to $R^{LL}_{sb,sd}$, $3$ of them are $\Delta^{LR}_{ij}$
with $ij= 13,23,33$ which contribute to $R^{RL}_{tb,td}$ and
$R^{LR}_{st}$. The other parameters  are $\tan\beta$, $M_2$, $\mu$
in chargino mass matrix, a common up-type squark mass $\tilde{m}$
and the mass of the (light) right-handed stop $m_{\tilde{t}_R}$.
We scan the parameter space in the region $4 \le \tan\beta \le
50$, $250 {\rm GeV} \le \tilde{m} \le 500 {\rm GeV}$, $90 {\rm
GeV} \le m_{\tilde{t}_R} \le 200 {\rm GeV}$, $100 {\rm GeV} \le
M_2 \le 500 {\rm GeV}$, $-500 {\rm GeV} \le \mu \le 500 {\rm
GeV}$. All the 9 $\Delta$'s are normalized by $\tilde{m}^2$
varying in the range between $-1$ and $1$. We find in the
calculation that a common feature is that Br($b \rightarrow s s
\bar{d}$) does not depend sensitively on $\tan\beta$, but rather
strongly on the mass parameters $\tilde{m}$, $m_{\tilde{t}_R}$ and
$m_{\tilde{\chi}^{\pm}_1}$, the lighter chargno mass .

\begin{figure}
\epsfxsize=16cm \epsfysize=16cm \centerline{\epsffile{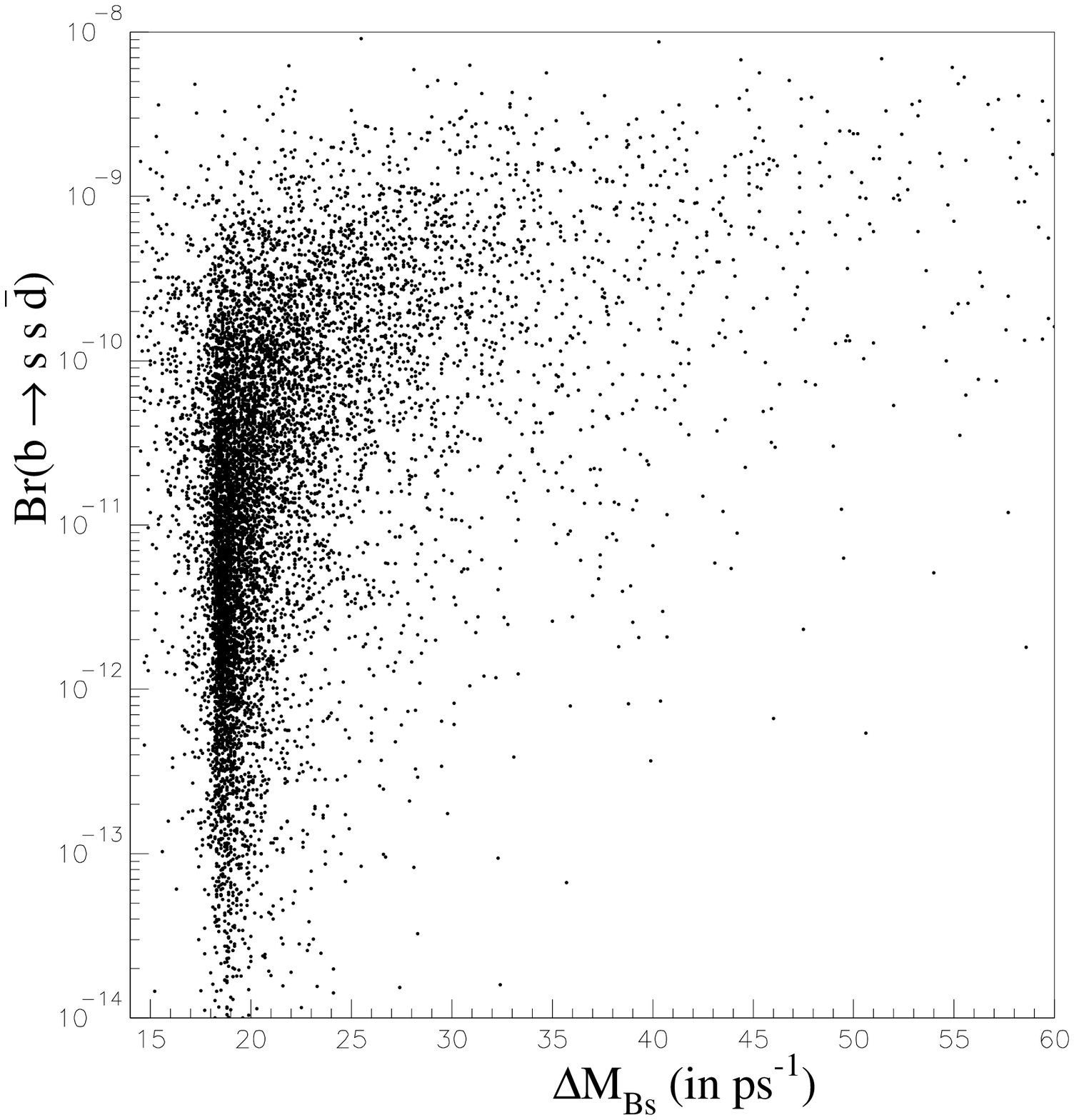}}
\caption{
Br($b \rightarrow s s \bar{d}$) {\it vs} $\Delta M_{B_s}$
with the contribution from $\Delta^{LR}_{13,23,33}$.
}
\label{scanlr}
\end{figure}

We find that the contribution to Br($b \rightarrow s s \bar{d}$)
from $\Delta^{LL}_{ij}$'s (with $ij=11,12,13,22,23,33$) and thus
from $R^{LL}_{sb,sd}$  is always below $10^{-11}$ if we use the
experimental constraints mentioned above.  The dominant
contribution to Br($b \rightarrow s s \bar{d}$) comes from
$\Delta^{LR}_{13,23}$, while $\Delta^{LR}_{33}$ is less important.
We  give in {\bf Tab}.~\ref{table} two representative points with
non-vanished $\Delta^{LR,LL}_{23}$ to show the main contribution
of $\Delta^{LR}_{ij}$.
%has only small effect with Br($b \rightarrow s
%s \bar{d}$) under $10^{-11}$.
We show  the $\Delta^{LR}_{13,23,33}$ contribution in {\bf
Fig}.~\ref{scanlr}. From {\bf Fig}.~\ref{scanlr}, Br($b
\rightarrow s s \bar{d}$) can be as large as $10^{-9}$. In some
region, with the increasing of Br($b \rightarrow s s \bar{d}$),
$\Delta M_{B_s}$ can be as large as $60{\rm ps}^{-1}$.

\begin{table}[!htb]
\begin{center}
\caption{Two points of the scan with $\tan\beta=10$. The other
$\Delta$'s which are not presented in the table are taken as zero.
} \label{table}
\begin{tabular}{rrrrr}
$\Delta$ ($\tilde{m}^2$) & $\tilde{m}$, $m_{\tilde{t}_R}$ (GeV) &
$m_{\tilde{\chi}^\pm_{1,2}}$ (GeV) & Br($b \rightarrow s s \bar{d}$) &
$\Delta M_{B_s}$ (${\rm ps}^{-1}$) \\
\hline
$\Delta^{LR}_{23} = -0.79$ & $493,124$ & $149,273$ &
$2.4 \times 10^{-10}$ & $27$\\
\hline
$\Delta^{LL}_{23} = -0.46$ & $322,103$ & $176,500$ &
$1.7 \times 10^{-12}$ & $19$\\
\hline
\end{tabular}
\end{center}
\end{table}

In conclusion, we have studied the  chargino contribution to
the rare decay $b \rightarrow s s \bar{d}$. We find that the
dominant contribution is from $\Delta^{LR}_{13,23}$ and the
branching ratio can amount to $10^{-9}$. This may be an important
source of $b \rightarrow s s \bar{d}$ in the MSSM.
% in some parameter region,
%and they also have large effects on mass difference of
%$\bar{B}^0_s - B^0_s$ system $\Delta M_{B_s}$, which can reach $60
%{\rm ps}^{-1}$. While the contribution from
%$\Delta^{LL}_{11,12,13,22,23,33}$ and $\Delta^{LR}_{33}$ are
%always small to Br($b \rightarrow s s \bar{d}$) under the level of
%$10^{-11}$.

We thank P. Singer for pointing out several mistakes.
The work of DXZ is supported in part by the National Natural
Science Foundation of China (NSFC) under the grant No. 90103014
and No. 10205001, and by the Ministry of Education of China.
And XHW's work is supported in part by the China Postdoctoral
Science Foundation.

\newpage
\begin{thebibliography}{99}
\bibitem{hlsz}  K.~Huitu, C.~D.~Lu, P.~Singer and D.~X.~Zhang,
%``Searching for new physics in b $\to$ s s anti-d decays,''
Phys.\ Rev.\ Lett. {\bf 81}, 4313 (1998).
\bibitem{hlsz2} K.~Huitu, C.~D.~Lu, P.~Singer and D.~X.~Zhang,
%``b $\to$ s s anti-d decay in two Higgs doublet models,''
Phys.\ Lett.\ B {\bf 445}, 394 (1999) [arXiv:hep-ph/9812253].
%%CITATION = HEP-PH 9812253;%%
\bibitem{data}G.~Abbiendi {\it et al.}  [OPAL Collaboration],
%``Search for new physics in rare B decays,''
Phys.\ Lett.\ B {\bf 476}, 233 (2000) [arXiv:hep-ex/0002008];
%%CITATION = HEP-EX 0002008;%%
J.~Damet, P.~Eerola, A.~Manara and S.~E.~Nooij,
%``Searching for physics beyond the standard model in the decay B+ $\to$ K+  K+ pi-,''
Eur.\ Phys.\ J.\ directC {\bf 3}, 7 (2001) [arXiv:hep-ex/0012057];
%%CITATION = HEP-EX 0012057;%%
A.~Garmash {\it et al.}  [Belle Collaboration],
%``Study of three-body charmless B decays,''
Phys.\ Rev.\ D {\bf 65}, 092005 (2002) [arXiv:hep-ex/0201007];
%%CITATION = HEP-EX 0201007;%%
A.~Garmash {\it et al.}  [Belle Collaboration],
%``Study of B meson decays to three-body charmless hadronic final states,''
arXiv:hep-ex/0307082.
%%CITATION = HEP-EX 0307082;%%
\bibitem{singer}S.~Fajfer and P.~Singer,
%``Constraints on heavy Z' couplings from Delta(S) = 2 B- $\to$ K- K- pi+  decay,''
Phys.\ Rev.\ D {\bf 65}, 017301 (2002) [arXiv:hep-ph/0110233].;
%%CITATION = HEP-PH 0110233;%%
S.~Fajfer and P.~Singer,
%``Search for new physics in Delta(S) = 2 two-body (VV, PP, VP) decays of  the B- meson,''
Phys.\ Rev.\ D {\bf 62}, 117702 (2000) [arXiv:hep-ph/0007132];
%%CITATION = HEP-PH 0007132;%%
Z.~j.~Xiao, W.~j.~Li, L.~b.~Guo and G.~g.~Lu,
%``Charmless decays B $\to$ P P, P V, and effects of new strong and  electroweak penguins in topcolor-assisted technicolor model,''
Eur.\ Phys.\ J.\ C {\bf 18}, 681 (2001) [arXiv:hep-ph/0011175];
%%CITATION = HEP-PH 0011175;%%
B.~Dutta, C.~S.~Kim and S.~Oh,
%``Charmless non-leptonic B decays and R-parity violating supersymmetry,''
Phys.\ Lett.\ B {\bf 535}, 249 (2002) [arXiv:hep-ph/0202019];
%%CITATION = HEP-PH 0202019;%%
E.~J.~Chun and J.~S.~Lee,
%``Wrong-sign kaons in B decays and new physics,''
arXiv:hep-ph/0307108;
%%CITATION = HEP-PH 0307108;%%
J.~P.~Saha and A.~Kundu,
%``Reevaluating bounds on flavor-changing neutral current parameters in R-parity conserving and R-parity violating supersymmetry,''
arXiv:hep-ph/0307259.
%%CITATION = HEP-PH 0307259;%%
\bibitem{buras98}A.~J.~Buras, A.~Romanino, L.~Silvestrini,
Nucl. Phys. {\bf B520} (1998) 3.
\bibitem{sto}A.~Stocchi, Nucl. Phys. Proc. Suppl. {\bf 117} (2003) 145.
\bibitem{pdg}K.~Hagiwara {\it et al.}, Particle Data Group,
Phys. Rev. {\bf D66} (2002) 010001.
\end{thebibliography}
\end{document}